\documentstyle[prd,tighten,aps]{revtex}
\begin{document}
\draft

\twocolumn[\hsize\textwidth\columnwidth\hsize\csname
@twocolumnfalse\endcsname
\renewcommand{\theequation}{\thesection . \arabic{equation} }
\title{\bf Quantum time machine}

\author{Pedro F. Gonz\'alez-D\'{\i}az}
\address{Centro de F\'{\i}sica ``Miguel Catal\'an'',
Instituto de Matem\'aticas y F\'{\i}sica Fundamental,\\
Consejo Superior de Investigaciones Cient\'{\i}ficas,
Serrano 121, 28006 Madrid (SPAIN)}
\date{December 6, 1997}

\maketitle

\begin{abstract}
The continuation of Misner space into the Euclidean region
is seen to imply the topological restriction that the
period of the closed spatial direction becomes time-dependent.
This restriction results in a modified Lorentzian Misner
space in which the renormalized stress-energy tensor for
quantized complex massless scalar fields becomes regular
everywhere, even on the chronology horizon. A
quantum-mechanically stable time machine
with just the sub-microscopic size may then be
constructed out of the modified Misner space, for which
the semiclassical Hawking's chronology protection conjecture
is no longer an obstruction.
\end{abstract}

\pacs{PACS number(s): 04.20.Gz, 04.62.+v }

\vskip2pc]

\renewcommand{\theequation}{\arabic{section}.\arabic{equation}}

\section{\bf Introduction}

After the seminal papers by Morris, Thorne and Yurtsever [1],
the notion of a time machine has jumped from the pages of
science fiction books to those of scientific journals,
giving rise to a recent influx of papers [2-7] and
books [8,9] on the subject. Several potentially useful
models for time machines have since been proposed, including
the wormhole of Thorne et al. [1], the Gott's couple of
cosmic strings set at relative motion [2], its Grant's
extension [3], the Politzer's machine [4], Roman rings [5]
and, ultimately, a time machine based on the so-called
ringholes [6].

Apart of all those difficulties related with potential
violation of classical causality that would arise while
travelling backward in time [8], all of these models
show at least one of the two following shortcomings:
an induced violation of the classical energy conditions
for the matter allowing closed timelike curves (CTC's) [9]
and, more importantly, the quantum instabilities
originating from the divergences of the stress-energy tensor
of vacuum fluctuations on the chronology horizon [8] (that
is the onset of the nonchronal region where CTC's develop).

Violation of energy conditions are generally produced by
the large inward pressure required on the tunnels and
leads to the emergence of regions where matter has
negative-energy density. This could no longer be a real
problem as negative energy has become commonplace in
quantum and semiclassical gravity [10,11]. More severe
is the difficulty related with the need of having a
tachyonic center of mass in parallel-moving cosmic
strings which are subject, moreover, to unphysical
boundary conditions [12].

By far the acutest problem with most of the proposed time
machines is the emergence of the stress-energy divergences
of the quantum vacuum fluctuations. This makes instable
the spacetime construct and has actually led to the
formulation of a chronology protection conjecture [13],
according to which the laws of physics should prevent
time travel to occur. So far, in spite of some attempts
intended to violate it [14,15], this conjecture has survived
rather forcefully [16]. However, the realm where chronology
protection holds is semiclassical physics, as it is for
all hitherto proposed time machines. Actually, because
spacetime foam [17] must entail strong violations of
causal locality everywhere, one would expect the
Hawking's conjecture to be inapplicable in the framework
of quantum gravity proper, and that the divergences of
stress-energy tensor due to quantum vacuum fluctuations
merely are artifact from inappropriate combination of
quantum matter fields with classical spacetime devices.

In this paper we consider an attempt to get rid of these
divergences in a nonchronal spacetime constructed out
from Misner space [18], where the thermal
properties of Euclidean gravity are taken into account.
The distinguishing feature of the considered spacetime
with respect to usual Misner space
is that the period of the closed spatial direction
becomes now time-dependent, and the nonchronal region
can only be acceded on if the time coordinate is
quantized according to a pattern which is the same as
that for the quantization being currently assumed [19]
for the event horizon of black holes. We have checked that
the stress-energy tensor for quantum vacuum fluctuations in
this spacetime is regular everywhere, but only time
machines with sub-microscopic size can be constructed
out of it.

The paper is organized as follows. In Sec. II we consider
the continuation of Misner space into the Euclidean sector
where the periodicity of the distinct spacetime directions
are fixed and then continued back to the Lorentzian region,
resulting in a modified structure for Misner space. The
Hadamard function and hence the expectation value of
the stress-energy tensor for quantum vaccum fluctuations
are evaluated in Sec. III in the case of a complex massless
scalar field propagating in the modified Misner space. It
is checked that these quantities are regular everywhere.
We conclude in Sec. IV, where the concept of a quantized
time machine is also discussed.
Throughout the paper units are used such that
$c=\hbar=G=1$.

\section{\bf The modified Misner space}
\setcounter{equation}{0}

It has been stressed [20]
that Misner space encompasses many remarkable
pathologies in the structure of spacetime, including those of
wormholes and ringholes and other spacetimes that
contain potentially interesting time machines. Therefore,
the spacetime structure and topology of the Misner space
itself remain being subjects which focuse current attention. It
is known that the metric of Misner space can be written as [21]
\begin{equation}
ds^2=-dt^2+t^2(dx^1)^2+(dx^2)^2+(dx^3)^2,
\end{equation}
where $0<t<\infty$, $0\leq x^2,x^3\leq\infty$, and $0\leq x^1\leq
2\pi$. With respect to Minkowski space, the fact that $x^1$ is
periodic is the unique distinguishing property originating
from its topology $S^1\times R^3$. Otherwise, they are
identical. This can be seen by introducing the covering space
through the coordinate transformations
\begin{equation}
y^0=t\cosh(x^1),\;\; y^0=t\sinh(x^1),\;\; y^2=x^2,\;\; y^3=x^3 ,
\end{equation}
where the metric becomes
\begin{equation}
ds^2=-(dy^0)^2+(dy^1)^2+(dy^2)^2+(dy^3)^2.
\end{equation}

If the period associated with direction $x^1$ is $a$, then the
topology of the Misner space implies the identification of
points on the covering space given by

\[(y^0,y^1,y^2,y^3)\leftrightarrow
(y^0\cosh(na)\]
\begin{equation}
+y^1\sinh(na),y^0\sinh(na)+y^1\cosh(na),y^2,y^3),
\end{equation}
which corresponds to the identifications
$(t,x^1,x^2,x^3)\leftrightarrow (t,x^1+na,x^2,x^3)$ in Misner
coordinates.

Let us now consider the Euclidean continuation of metric (2.1).
Doing that will allow us to investigate how periodical properties
of the distinct space directions in the Euclidean sector
transform when they are rotated back to the Lorentzian region.
Metric (2.1) becomes definite positive if we introduce the
rotation
\begin{equation}
t=i\tau,\;\;\; x^1=i\chi .
\end{equation}
Then,
\begin{equation}
ds^2=d\tau^2+\tau^2(d\chi)^2+(dx^2)^2+(dx^3)^2.
\end{equation}

An interesting property
of the Euclideanized Misner space is that its
covering space preserves the Lorentzian signature of the
Minkowskian covering for the original Misner space. From
(2.2) and (2.5) we have now the coordinate transformations:
\begin{equation}
\tilde{y}^0=i\tau\cos\chi,\;\; \tilde{y}^1=-\tau\sin\chi,
\;\; \tilde{y}^2=x^2, \;\; \tilde{y}^3=x^3 ,
\end{equation}
so that
\begin{equation}
ds^2=-(d\tilde{y}^0)^2+(d\tilde{y}^1)^2+
(d\tilde{y}^2)^2+(d\tilde{y}^3)^2.
\end{equation}
However, the topology that corresponds to Euclideanized Misner
space is different from that is associated with (2.3). For
(2.8), the topology is $S^1\times S^1\times R^2$. This can be
seen by noting that in the coordinates of metric (2.1), this
metric has an apparent singularity at $t=0$. One can still
extend metric (2.1) beyond $t=0$ by using new coordinates
defined by [21]
\begin{equation}
T=t^2,\;\;\; V=\ln t + x^1 ,
\end{equation}
with which metric (2.1) transforms into
\begin{equation}
ds^2=-dVdT+TdV^2 .
\end{equation}

In order to investigate the periodic properties of the directions
in the Euclideanized Misner space, one can introduce now the
extended coordinate definitions of the Lorentzian Misner
extended metric
\begin{equation}
u+w=T,\;\;\; u-w=V,
\end{equation}
with which (2.10) becomes
\begin{equation}
ds^2=-du^2+dw^2+(u+w)(du-dw)^2
\end{equation}
and
\begin{equation}
u^2-w^2=TV, \;\; \frac{u-w}{u+w}=\frac{V}{T} .
\end{equation}

Metric (2.12) will become positive definite provided we allow the
continuation $u=i\zeta$; hence we have $\zeta^2+w^2=-TV$ and,
from (2.9), it can be seen that the section on which $\zeta$
and $w$ are both real will correspond to $0\leq t\leq 1$,
$-2\pi\leq x^1\leq 0$. That section defines a Misner instanton
describing transitions on the chronal region. Using then (2.5)
and the second of (2.13) we finally obtain:
\[\exp\left(-\frac{2i\zeta w}{\zeta^2+w^2}\right)=\]
\begin{equation}
\tau^{\tau^{-2}}\exp\left(\frac{w^2-\zeta^2}{w^2+\zeta^2}\right)
\exp\left[i\left(\frac{\pi/2+\chi}{\tau^2}\right)\right].
\end{equation}
It follows that on the Euclidean sector, both $\tau$ and $\chi$
are periodic, with respective periods
\[\Pi_{\tau}=\frac{1}{2},\;\; \Pi_{\chi}=2\pi\tau^2=g(\tau) .\]
Therefore, the topology of the Euclidean Misner space is
$S^1\times S^1\times R^2$, and one should expect that an
observer in Misner space would detect a thermal bath at
temperature $T_M=2$.

Rotating back to the Lorentzian sector (i.e. taking $\tau=-it$
and $\chi=-ix^1$) we see that $t$ becomes no longer periodic,
though $x^1$ still keeps a periodic character, with a period
which should be given by
\begin{equation}
a=|g(t)|=2\pi t^2 .
\end{equation}
Thus, the continuation to the Euclidean section of Misner space
shows that the Lorentzian Misner space itself must be modified in
that the period of coordinate $x^1$ should no longer be an arbitrary
constant, but it depends on time $t$ according to (2.15). This
change manifests in that the extension of the region covered by
metric (2.1) accross the chronology boundary $t=0$ into a new region
that contains CTC's becomes modified in that
the twisted null geodesics [21] spiral round and round now each time
with a larger frequency as they approach $t=0$,
to finally diverge at
this point. On the Minkowskian covering space this modification
translates as a concentration onto a single point of all the
identified points on the surface $\sigma=(y^0)^2-(y^1)^2=0$
of the left and right chronology horizons.
Actually, the need of making period $a$
proportional to $t^2$ arises from the kind of
instantonic quantization
implied by continuing into the Euclidean sector, so that the
modified Misner space can be regarded as essentially
semiclassical.

Removal of the resulting
singularity at $t=0$ can only be achieved if we let time $t$ to be
quantized so that either (i) the point at $t=0$
is replaced by a minimum
nonzero throat (i.e. by transforming $t$ according to
$t^2\rightarrow t^2+R_0^2$, with $R_0^2$ a given small
constant which defines the throat, so
restoring the usual Misner space situation where twisted
null geodesics terminated at $t=0$, leading to the
existence of two inequivalent geodesically incomplete
analytic extensions of (2.1), $V_{\pm}=\pm\ln t+x^1$,
which are locally inextendible [21], or (ii), if we want
to establish full equivalence between the two possible
analytic extensions $V_{\pm}$ (so that all null geodesics
do not terminate at $t=0$ but are extendible beyond it),
then time $t$ must be quantized in such a way that though
the two extensions cover the whole space only some particular
values of time are allowed.

A simplest ansatz that implements this requirement is
\[t^2=(n+\beta)C ,\]
where $n=0,1,2,3,...$, $\beta$ is an arbitrary parameter of
order unity, and $C$ defines again a
throat at $n=0$.
In the next section we shall see that it is such an ansatz
what is
actually required if we want a scalar field propagating
in the modified Misner space to be expandible in an
orthonormal basis with time-independent frequency.

\section{\bf The Hadamard function}
\setcounter{equation}{0}

In what follows we shall consider the propagation of a complex
scalar field $\phi$ in the modified Misner space discussed
in Sec. II. For the sake of generality we will require $\phi$
to satisfy the automorphic condition [22]
($X\equiv t,x^1,x^2,x^3$)
\begin{equation}
\phi(\gamma X)=e^{2\pi i\alpha}\phi(X), \;\;
0\leq\alpha\leq\frac{1}{2},
\end{equation}
with $\alpha$ the automorphic parameter and $\gamma$ representing
the symmetry (periodicity identifications) transformation of
the modified Misner space; i.e.: the transformation will be
taken to satisfy the periodicity condition for a period
$2\pi t^2$. We shall follow the analysis carried out by
Sushkov [14] and look at complex scalar fields that obey the
field equation
\[\Box\phi=\Box\bar{\phi}=0.\]
On the covering space coordinates, this equation admits the
general positive-frequency solution:
\[\phi(y^0,y^1,y^2,y^3)=\]
\begin{equation}
\int dk_1 dk_2 dk_3 \frac{A(k_0,k_1,k_2,k_3)}{4\pi^{\frac{3}{2}}k_0}
e^{i(-k_0 y^0+k_1 y^1+k_2 y^2+k_3 y^3)} ,
\end{equation}
where $k_0=\sqrt{k_1^2+k_2^2+k_3^2}$ and $A(k_0,k_1,k_2,k_3)$ is
a given spectral function.

The demand that $\phi$ satisfies condition (3.1) amounts to the
functional relation
\[\phi (y^0\cosh(2\pi t^2)+y^1\sinh(2\pi t^2),
y^0\sinh(2\pi t^2)\]
\begin{equation}
+y^1\cosh(2\pi t^2),y^2,y^3)=e^{2\pi i\alpha}\phi(y^0,y^1,y^2,y^3),
\end{equation}
with which automorphicity for the solutions will hold provided
\[A(k_0\cosh(2\pi t^2)-k_1\sinh(2\pi t^2),
-k_0\sinh(2\pi t^2)\]
\begin{equation}
+k_1\cosh(2\pi t^2),k_2,k_3)=e^{2\pi i\alpha}A(k_0,k_1,k_2,k_3).
\end{equation}
This admits a general solution which is formally the same as
that obtained by Sushkov [14], i.e.
\begin{equation}
A(k_0,k_1,k_2,k_3)=\sum_{n=0}^{\infty}C_n(k_2,k_3)(k_0-k_1)^{i\nu},
\end{equation}
(where again the $C_n(k_2,k_3)$ are arbitrary functions of $k_2$
and $k_3$), but in which the frequency $\nu$ would in principle
have explicit dependence on time:
\begin{equation}
\nu=-\frac{n+\alpha}{t^2}.
\end{equation}

Demand of time-independence for frequency $\nu$ implies quantization
of time $t$ according to the law
\begin{equation}
t^2=(n+\alpha)t_0^2,
\end{equation}
where $t_0^2$ is an arbitrary constant. This quantization condition
for time $t$ is exactly of the same form as that of the ansatz
used in Sec. II in order to avoid the singularity at $t=0$,
and leads finally to a frequancy given by $\nu=-t_0^{-2}$.

Inserting (3.5) into (3.2) and carrying first out the integration
over $k_1$ and then transforming to Misner coordinates [14],
we obtain
finally for the automorphic field
\[\phi(t,x^1,x^2,x^3)=\]
\begin{equation}
\sum_{n=0}^{\infty}\int dk_2\int dk_3
\tilde{C}_n(k_2,k_3)\varphi(t,x^1,x^2,x^3),
\end{equation}
where
\begin{equation}
\varphi(t,x^1,x^2,x^3)=D(n)H_{-it_0^{-2}}^{(2)}(kt)
e^{i(-t_0^{-2}x^1+k_2 x^2+k_3 x^3)} ,
\end{equation}
with $H_{i\nu}^{(2)}(kt)$ the Bessel function of the third kind,
$D(n)$ a normalizing coefficient, and $k=\sqrt{k_2^2+k_3^2}$.
Adopting the definition of scalar product of Ref. 14, one can now
show that the solutions (3.9) form an orthonormal basis in a
Hilbert space provided we choose
\begin{equation}
D(n)=\frac{\sqrt{2}}{8\pi te^{\frac{\pi(n+\alpha)}{2t^{2}}}}
=\frac{\sqrt{2}}{8\pi\sqrt{n+\alpha}t_0 e^{\frac{\pi}{2t_0^{2}}}} .
\end{equation}

Noting that for two generally different sets of Misner
coordinates ($X\equiv\{x^i\}$ and $\tilde{X}\equiv\{\tilde{x}^i\}$,
$i=0,1,2,3$) the corresponding arbitrary frequencies $\nu$
and $\tilde{\nu}$ are the same, we can finally calculate
the renormalized Hadamard function. This will be taken
to be given by ($\mu,\nu=0,1,2,3$, $\beta,\gamma=1,2,3$)
\[G_{ren}^{(1)}(X^{\beta},\tilde{X}^{\gamma})=
\left.\sum_{\tilde{n}=0}^{\infty}G_{ren}^{(1)}(X^{\mu},\tilde{X}^{\nu})
\right|_{t=\sqrt{n+\alpha}t_0,\tilde{t}=\sqrt{\tilde{n}+\alpha}t_0}\]
\[=\left(\sum_{n=0}^{\infty}\sum_{\tilde{n}=0}^{\infty}\int\int dk_2 dk_3
\left[\varphi(X^{\mu})\bar{\varphi}(\tilde{X}^{\nu})+
\varphi(\tilde{X^{\mu}})\bar{\varphi}(X^{\nu})\right]\right.\]
\[\left.\left.-G_0^{(1)}(X^{\mu},\tilde{X}^{\nu})\right)
\right|_{t=\sqrt{n+\alpha}t_0,\tilde{t}=\sqrt{\tilde{n}+\alpha}t_0}\]
\[=\left.\frac{1}{4\pi^2}\left(\frac{1}{\sigma(X^{\mu},\tilde{X}^{\nu})}-
\frac{1}{\sigma_0(X^{\mu},\tilde{X}^{\nu})}\right)
\right|_{t=\tilde{t}=\sqrt{\alpha}t_0}\]
\begin{equation}
+\left.\frac{1}{\pi^2}\sum_{n=1}^{\infty}\sum_{\tilde{n}=1}^{\infty}
\frac{\cosh[n(x^1-\tilde{x}^1)]\sin(n\arccos\chi)}
{t\tilde{t}\sqrt{1-\chi^2}}\Psi_{n}(\alpha,t)
\right|_{A} ,
\end{equation}
where the subscript $A$ means evaluation at
\[A\{t=\sqrt{n+\alpha}t_0,\tilde{t}=\sqrt{\tilde{n}+\alpha}t_0\} ,\]
$G_{ren}^{(1)}(X^{\mu},\tilde{X}^{\nu})$ is the Hadamard
function obtained in Ref. 14, with $\Psi_{n}(\alpha,t)$ given
now by
\begin{equation}
\Psi_{n}(\alpha,t)=
\frac{e^{2\pi nt^2}\cos(2\pi\alpha)-1}{e^{4\pi nt^2}-
2e^{2\pi nt^2}\cos(2\pi\alpha)+1},
\end{equation}
and
\[\sigma(X^{\mu},\tilde{X}^{\nu})=\frac{1}{2}\{-(t-\tilde{t})^2\]
\begin{equation}
+2t\tilde{t}[\cosh(x^1-\tilde{x}^1)-1]
+(x^2-\tilde{x}^2)^2
+(x^3-\tilde{x}^3)^2\} ,
\end{equation}
\[\sigma_0(X^{\mu},\tilde{X}^{\nu})=\frac{1}{2}\{-(t-\tilde{t})^2\]
\begin{equation}
+t^2(x^1-\tilde{x}^1)+(x^2-\tilde{x}^2)^2
+(x^3-\tilde{x}^3)^2\} .
\end{equation}
The final result of this calculation is:
\[G_{ren}^{(1)}(X^{\beta},\tilde{X}^{\gamma})
=\frac{1}{4\pi^2}\left(\frac{1}{\sigma(X^{\beta},\tilde{X}^{\gamma})}-
\frac{1}{\sigma_0(X^{\beta},\tilde{X}^{\gamma})}\right)\]
\begin{equation}
+\frac{1}{\pi^2}\sum_{n=1}^{\infty}\sum_{\tilde{n}=1}^{\infty}
\frac{\cosh[n(x^1-\tilde{x}^1)]\sin(n\arccos\chi)}
{\sqrt{n+\alpha}\sqrt{\tilde{n}+\alpha}t_0^2\sqrt{1-\chi^2}}\Psi_{n}(\alpha)
 ,
 \end{equation}
where
\begin{equation}
\Psi_{n}(\alpha)=
\frac{e^{2\pi n(n+\alpha)t_0^2}\cos(2\pi\alpha)-1}{e^{4\pi n(n+\alpha)t_0^2}-
2e^{2\pi n(n+\alpha)t_0^2}\cos(2\pi\alpha)+1},
\end{equation}
\[\sigma(X^{\beta},\tilde{X}^{\gamma})=\frac{1}{2}\{
2\alpha t_0^2[\cosh(x^1-\tilde{x}^1)-1]\]
\begin{equation}
+(x^2-\tilde{x}^2)^2
+(x^3-\tilde{x}^3)^2\} ,
\end{equation}
\[\sigma_0(X^{\beta},\tilde{X}^{\gamma})=\frac{1}{2}\{
\alpha t_0^2(x^1-\tilde{x}^1)+(x^2-\tilde{x}^2)^2\]
\begin{equation}
+(x^3-\tilde{x}^3)^2\} .
\end{equation}

It follows that the
value of $G_{ren}^{(1)}(X^{\beta},\tilde{X}^{\gamma})$ remains finite
on the whole modified Misner space for any value of $\alpha$,
even $\alpha=0$, provided the arbitrary constant $t_0$ is
nonzero.
This result inmediately
reflects in the remarkable implication that the
renormalized vacuum expectation values of the scalar
field squared, $\langle\phi^2\rangle=\langle 0|\phi^2|0\rangle$,
and stress-energy tensor,
$\langle T_{\mu\nu}\rangle=\langle 0|T_{\mu\nu}|0\rangle$,
remain both finite as well on the nonsingular
modified Misner space. These quantities are evaluated as
follows:
\begin{equation}
\langle\phi^2\rangle=
\lim_{\tilde{X}\rightarrow X}G_{ren}^{(1)}(X^{\beta},\tilde{X}^{\gamma})
=\frac{1}{\pi^2 t_0^2}\sum_{n=1}^{\infty}
\frac{n\Psi(\alpha)}{n+\alpha} ,
\end{equation}
\[\langle T_{\beta\gamma}\rangle=\lim_{\tilde{X}\rightarrow X}
\{[(1-\xi)\nabla_{\mu}\tilde{\nabla}_{\nu}
+(2\xi-\frac{1}{2})g_{\mu\nu}\nabla_{\alpha}\tilde{\nabla}^{\alpha}\]
\[-2\xi\nabla_{\mu}\nabla_{\nu}+\frac{1}{2}\xi g_{\mu\nu}\Box]
\left.G_{ren}^{(1)}(X^{\mu},\tilde{X}^{\nu})
\right|_{t=\sqrt{n+\alpha}t_0,\tilde{t}=\sqrt{\tilde{n}+\alpha}t_0}\} \]
\begin{equation}
=\frac{1}{\pi^2 t_0^4}\sum_{n=1}^{\infty}
\frac{diag(L,3L,M,M)}{(n+\alpha)^2}+\frac{1}{\pi^2}F(\alpha,t_0) ,
\end{equation}
where $F(\alpha,t_0)$ is a complicate function of $\alpha$ and
$\xi$ arising from the time derivatives of the function
$\Psi_n(\alpha,t)$, which is regular
for all values of $\alpha$ and $\xi$;
$\xi$ is the coupling parameter for the scalar field
($\xi=0$ for minimal coupling and $\xi=\frac{1}{6}$ for
conformal coupling) and $L$ and $M$ are as given in Ref. 14,
but with $\Psi_n(\alpha,a)$ replaced for $\Psi_n(\alpha)$,
as given in (3.16), that is
\[L=\frac{1}{\pi^2}\sum_{n=1}^{\infty}
\left(\frac{n-n^3}{3}-\frac{3\xi n}{2}\right)\Psi_n(\alpha)\]
\[M=\frac{1}{\pi^2}\sum_{n=1}^{\infty}
\left(\frac{2n+n^3}{3}-\frac{9\xi n}{2}\right)\Psi_n(\alpha) .\]
For the particular
case of non-twisted fields, $\alpha=0$, expression
(3.12) reduces to
\[\Psi(t_0)=\left(e^{2\pi n^2 t_0^2}-1\right)^{-1},\]
so that the quantities $\langle\phi^2\rangle$ and
$\langle T_{\beta\gamma}\rangle$ are finite even when
$\alpha=0$.

Thus, all the divergences coming from quantum vacuum
fluctuations that take place on the chronology horizon
of the classical Misner space whose periodic direction has
constant period are smoothed out on the modified, no longer
classical Misner space where the period along the periodic
direction is time-dependent, for which case the expectation
values of the scalar field squared and stress-energy tensor
are regular everywhere for any value of $\alpha$ and $\xi$,
provided $t_0\neq 0$.

\section{\bf Conclusions: The quantum time machine}

In this paper we have considered the change of topology implied
by the continuation from Lorentzian Misner space into its
Euclidean counterpart; i.e.: from $S^1\times R^3$ to
$S^1\times S^1\times R^2$. It was seen that the spatial
direction which is also periodic in the Euclidean continuation
has a period that depends on the Euclidean time and rotates
back to an also time-dependent period on the Lorentzian sector.
This induces a change in the structure of the Misner space,
giving rise to a true singularity at $t=0$, instead of the
apparent singularity on the chronology horizon. A chronology
horizon with nonzero width
can only be restored if time is quantized according
to a rule which parallels that assumed for the quantization
of the black hole event horizon [19].

We can show that the region $t<0$ of the modified Misner
space keeps its nonchronal character everywhere
only if $\alpha t_0^2<1$, that is in the quantum-gravity
regime. Consider the line $L_1$ defined by $t=-\omega x^1$,
$x^2=x^3=0$, which will be timelike everywhere provided
$2\pi\omega<1$, and the circle $L_0$ defined by
$t=\sqrt{\alpha}t_0>0$, $x^2=x^3=0$, which is also timelike
everywhere if $\sqrt{\alpha}t_0<1$. It is then easy to see
that on $L_1$, the point $Q$($x^1=\frac{\sqrt{\alpha}}{\omega}$)
precedes the point $P$($x^1=2\pi +\frac{\sqrt{\alpha}}{\omega})$,
but these two points are also on circle $L_0$ where $P$ precedes
$Q$, provided $\omega>0$. Therefore, there will be CTC's always
that the condition
\[1>2\pi\omega\geq\sqrt{\alpha}t_0 ,\]
holds, and hence the above conclusion follows.

The modified Misner space could be therefore transformed into a
time machine by simply replacing the identified flat planes
for identified spheres or tori, and allowing one of the
resulting spherical or toroidal mouths to move relative
to the respective other mouth. This would be equivalent
to extract two spheres or tori from three-dimensional
Euclidean space and identify the sphere or torus surfaces,
while they are set in relative motion, so when you enter
the surface of one you find yourself emerging from the
surface of the other. In Minkowski spacetime, the time
machine is obtained identifying the two world tubes swept
out by the approaching (or receding) spheres or tori,
with events at the same Lorentz time identified.
The result would be what one may call {\it quantum time
machine}. The chronology horizon of this construct is not
on just the single surface at $t=0$, but is spread
throughout a time strip with nonzero width, around the
surface at $t=0$.

We have also checked that the quantum
vacuum polarization has no catastrophic
effects anywhere in the modified Misner space, so quantum
time machines are quantum-mechanically stable. One would
moreover expect that when their size approach the Planck
length scale, these machines are spontaneously created
in the quantum-gravity framework of the spacetime foam [17].
We do not know the probability for the existence of
such machines, but the very definition of the foam requires
violation of causal locality everywhere, and hence one can
assume the existence of submicroscopic quantum time machines
as a pre-requisite for the existence of the spacetime foam
itself, and may therefore adscribe a probability of order
unity for the existence of time machines in the foam.

The possibility that a future technology [1],
or present natural
process taking place somewhere in the universe [23],
be able to
pull a quantum time machine with extremely large spacetime
curvature out from the quantum spacetime foam, and then grow
it up to macroscopic size [1], will largely
reside on the probability of existence that the macroscopic
machine may have. Such a probability may still be estimated
by instantonic techniques in the semiclassical approximation.
On the $\tau$-$\chi$ section of the instanton studied in
Sec. II, any boundary has topology $S^1\times S^1$ and so
is compact. Since the scalar curvature of that section vanishes,
the action can then be written only in terms of the surface
integrals corresponding to the fixed boundaries. For the
case being considered, this action is:
\[I_E\simeq\frac{1}{8\pi}\int_{\partial M}dx^1 (K-K^0)
=\frac{1}{8\pi}\int_{0}^{a}dx^1=\frac{1}{4},\]
where $K$ is the trace of the second fundamental form of
the boundary and $K^0$ the same for the boundary embedded
in flat space. Hence, the semiclassical probability,
$P\simeq e^{-I_{E}}$, for the spontaneous creation
of macroscopic
time machines would be expected to be of order unity too.

The chronology protection conjecture advanced by Hawking
[13] has an
essential semiclassical character. Therefore, the feature
that a quantum time machine has not quantum instabilities
does not necessarily imply violation of the conjecture,
but rather the
need for a new one which would state that fully quantized
laws of physics {\it require} the existence of stable time
machines, at least in the framework of quantum gravity,
since otherwise no spacetime foam could exist.  Whether
or not this conjecture would ultimately imply the possibility
of technologically growing one of these sub-microscopic
time machines up to a macroscopic size that remains stable
is a matter which
our present knowledge does not allow to decide on. Nevertheless,
if we keep in the semiclassical regime where no quantization
of time can be assumed, then the true singularity at $t=0$
of the modified Misner space will
prevent access to the nonchronal region and, in at least
this sense, the Hawking's conjecture would continue to hold.

\acknowledgements

\noindent For useful comments, the author thanks
L.J. Garay and G.A. Mena Marug n of IMAFF. This
research was supported by DGICYT under Research Projects No.
PB94-0107 and, partially, No. PB93-0139.


\begin{references}
\bibitem {1} M.J. Morris, K.S. Thorne and U. Yurtsever, Phys. Rev. Lett.\c{c}
61, 1446 (1988); M.S. Morris and K.S. Thorne, Am. J. Phys. 56, 395 (1988).
\bibitem {2} J. Gott, Phys. Rev. Lett. 66, 1126 (1991).
\bibitem {3} J.D.E. Grant, Phys. Rev. D47, 2388 (1993).
\bibitem {4} H.D. Politzer, Phys. Rev. D49, 3981 (1994).
\bibitem {5} T.A. Roman, Phys. Rev. D47, 1370 (1993).
\bibitem {6} P.F. Gonz\'alez-D\'{\i}az, Phys. Rev. D54, 6122 (1996).
\bibitem {7} B. Jensen and H.H. Soleng, Phys. Rev. D45, 3528 (1992).
\bibitem {8} K.S. Thorne, {\it Black Holes and Time Warps} (Norton,
New York, 1994).
\bibitem {9} M. Wisser, {\it Lorentzian Wormholes} (Am. Inst. Phys.
Press, Woodbury, 1996).
\bibitem {10} N.D. Birrell and P.C.W. Davies, {\it Quantum fields
in curved space} (Cambridge Univ. Press, Cambridge, England, 1982).
\bibitem {11} M.J. Spaarnay, Physica 24, 751 (1958);
D. Tabor and R.H.S. Winterton, Proc. Roy. Soc. Lond., A312,
435 (1969).
\bibitem {12} S. Deser and A.R. Steif, in: {\it Directions in
General Relativity. Vol. 1}, edited by B.L. Hu, M.P. Ryan Jr.,
and C.V. Vishveshwara (Cambridge Univ. Press, Cambridge, England,
1993).
\bibitem {13} S.W. Hawking, Phys. Rev. D46, 603 (1992).
\bibitem {14} S.V. Sushkov, Class. Quant. Grav. 14, 523 (1997).
\bibitem {15} S.V. Krasnikov, Phys. Rev. D54, 7322 (1996).
\bibitem {16} B.S. Kay, M.J. Radzikowski and R.M. Wald, Commun.
Math. Phys. 183, 533 (1997).
\bibitem {17} S.W. Hawking, Nucl. Phys. B144, 349 (1978).
\bibitem {18} C.W. Misner, in {\it Relativity Theory and Astrophysics I.
Relativity and Cosmology}, edited by J. Ehlers (American
Mathematical Society, Providence, RI, 1967).
\bibitem {19} J.D. Bekenstein, Lett. Nuovo Cimento 11, 467 (1974);
{\it Quantum Black Holes as Atoms}, gr-qc/9710076.
\bibitem {20} K.S. Thorne, in: {\it Directions in
General Relativity. Vol. 1}, edited by B.L. Hu, M.P. Ryan Jr.,
and C.V. Vishveshwara (Cambridge Univ. Press, Cambridge, England,
1993).
\bibitem {21} S.W. Hawking and G.F.R. Ellis, {\it The large
scale structure of space-time} (Cambridge Univ. Press,
Cambridge, England, 1973).
\bibitem {22} R. Banach and J.S. Dowker, J. Phys. A12,
2527 (1979).
\bibitem {23} P.F. Gonz lez-D¡az, Phys. Rev. D56, 6293 (1997).
\end{references}
\end{document}